\documentclass{aa}
\usepackage{graphicx}
\usepackage{natbib}
\usepackage{psfig}
\newcommand\nodata{...}
\newcommand\eg{{\it e.g.} }

\newcommand\etal{et~al.}

\newcommand\Lya{Ly$\alpha$}

\newcommand\OII{\ion{[O}{II]}~$\lambda$~3727}

\newcommand\Ha{H$\alpha$}

\newcommand\kms{\ifmmode {\rm\,km\,s^{-1}}\else${\rm\,km\,s^{-1}}$\fi}
\font\aipsfont = cmsy9 scaled\magstep1

\newcommand\aips {{\aipsfont AIPS}}
\def\spose#1{\hbox to 0pt{#1\hss}}
\newcommand\simlt{\mathrel{\spose{\lower 3pt\hbox{$\mathchar"218$}}
     \raise 2.0pt\hbox{$\mathchar"13C$}}}
\newcommand\simgt{\mathrel{\spose{\lower 3pt\hbox{$\mathchar"218$}}
     \raise 2.0pt\hbox{$\mathchar"13E$}}}
\begin{document}
\title{A multi-wavelength study of the proto-cluster\\ surrounding the $z$=4.1 radio galaxy TN~J1338$-$1942.\thanks{Based on observations obtained with the IRAM 30m, the Very Large Array, the James Clerk Maxwell Telescope, and the ESO Very Large Telescope at Paranal, Chile (programs LP167.A-0409, 69.B-0078 and 71.A-0495).}}
\titlerunning{Multi-wavelength study of a $z$=4.1 proto-cluster.}

\author{Carlos De Breuck\inst{1,2} \and Frank Bertoldi\inst{3} \and Chris Carilli\inst{4} \and Alain Omont\inst{2} \and  Bram Venemans\inst{5} \and Huub R\"ottgering\inst{5} \and Roderik Overzier\inst{5} \and Michiel Reuland\inst{5,6,7} \and George Miley\inst{5} \and Rob Ivison\inst{8} \and Wil van Breugel\inst{6}}

\offprints{Carlos De Breuck}
\institute{European Southern Observatory, Karl Schwarzschild Stra\ss e 2, D-85748 Garching, Germany \\ \email{cdebreuc@eso.org}
\and Institut d'Astrophysique de Paris, CNRS, 98bis Boulevard Arago, F-75014 Paris, France \\ \email{omont@iap.fr} 
\and Max Planck Institut f\"ur Radioastronomie, Auf dem H\"ugel 69, D-53121 Bonn, Germany \\ \email{bertoldi@mpifr-bonn.mpg.de} 
\and National Radio Astronomy Observatory, P.O. Box O, Socorro, NM 87801, USA \\  \email{ccarilli@nrao.edu} 
\and Sterrewacht Leiden, Postbus 9513, NL-2300 RA Leiden, The Netherlands\\ \email{rottgeri,venemans,overzier,miley@strw.leidenuniv.nl} 
\and IGPP/LLNL, L-413, 7000 East Ave, Livermore, CA 94550, USA\\ \email{mreuland,wil@igpp.ucllnl.org} 
\and Department of Physics, University of California, Davis, CA 95616, USA
\and Institute for Astronomy, University of Edinburgh, Royal Observatory, Blackford Hill, Edinburgh EH9 3HJ, United Kingdom\\ \email{rji@roe.ac.uk}}

\date{Received 2003 December 16; accepted 2004 May 14}

\abstract{
We present a 1.2~mm (250~GHz) map obtained with MAMBO on the IRAM 30m telescope of the central 25~arcmin$^2$ of the proto-cluster surrounding the $z$=4.1 radio galaxy TN~J1338$-$1942. The map reaches a 1$\sigma$ sensitivity of 0.6~mJy in the central area, increasing to 1.2~mJy at the edges. We detect 10 candidate mm sources, of which 8 are also detected in a deep VLA 1.4~GHz map and/or a VLT $R-$band image. Three sources have a flux density $S_{\rm 1.2 mm}>$4.0~mJy, representing a 7$\sigma$ overdensity compared to random field surveys, which predict only 1 such source in our map area. We obtained SCUBA/JCMT 850~$\mu$m and 450~$\mu$m photometry of six radio/optically identified MAMBO sources, confirming 5 of them with S/N$>$4. Radio-to-mm and mm-to-submm redshift estimators cannot put strong constraints on the redshifts of these MAMBO sources, but 9 of them are consistent within the uncertainties (mean $\Delta z$=+2.6) with $z$=4.1. One faint MAMBO source is possibly identified with an extremely red object ($R-K$=6.1) at a likely spectroscopic redshift $z$=1.18.\\
The four brightest MAMBO sources are all located north of the radio galaxy, while the densest area of companion \Lya\ excess and Lyman break galaxies is to the southeast. None of the 14 spectroscopically confirmed \Lya\ emitters in the MAMBO field are detected at 1.2~mm; their average 1.2~mm flux density is $\langle S_{\rm 1.2mm}\rangle$=0.25$\pm$0.24~mJy.
If the mm sources lie at $z$=4.1, none of them show excess \Lya\ emission in our narrow-band images. Both populations thus show no apparent overlap, possibly due to dust quenching the \Lya\ emission. If the mm sources are part of the proto-cluster, our results suggest that galaxies with star formation rates of a few 1000~M$_{\odot}$yr$^{-1}$ could be spread throughout the proto-cluster over projected scales of at least 2~Mpc.
 \\
}

\maketitle
%

\section{Introduction}

Our knowledge of large-scale structures at high redshifts has made significant progress in recent years with the detection of several overdense regions out to $z>4$ \citep[\eg][]{ste98,pen00b,ven02,shi03}.
The discovery of these proto-clusters with sizes of several Mpc was made possible by deep wide-field optical surveys, concentrating on a narrow redshift interval using the Lyman-break or narrow-band \Lya\ \citep{kurk00,ven02} or \Ha\ \citep{kurk04} excess techniques, followed by multi-slit spectroscopy on 8-10m telescopes. Similar surveys have also been done in random fields \citep[\eg][]{hu98,hu02,rho03,kod03,shi03}.

Hierarchical galaxy formation models \citep[\eg][]{kau99} predict that the best fields to search for such overdensities are those containing a massive galaxy. High redshift radio galaxies (HzRGs; $z>2$) are therefore ideal targets, as they are not only among the most massive galaxies known at high $z$ \citep[\eg][]{deb02,roc04}, but their lower redshift ($z \simgt 0.5$) counterparts are also located in cluster environments \citep[\eg][]{hill91,best00,best03}. This motivated an intensive search of \Lya\ emitting companions around six radio galaxies with $2.1<z<4.1$, leading to the discovery of overdensities of \Lya\ emitters of 5 to 15 compared with the random fields \citep{ven02,ven03}.

However, the \Lya\ excess technique detects only a fraction of the companion objects. For example, in the proto-cluster of Lyman break galaxies at $z$=3.09 in the SSA~22 region \citep[LBG;][]{ste98}, only 25\% of the LBGs have a sufficiently high \Lya\ equivalent width to be included in a \Lya\ excess selection \citep{ste00}. This incompleteness is also illustrated by the presence of a similar overdensity of \Ha\ emitters and extremely red objects in the proto-cluster surrounding the $z$=2.16 radio galaxy PKS~1138$-$262, while these galaxies are mostly not detected using the \Lya\ excess technique \citep{kurk04}. Deep {\it Chandra} observations of this field have also found an overdensity of X-ray sources, of which at least 2 are AGN within the proto-cluster \citep{pen02}. \citet{sma03b} also report the detection of four {\it Chandra} X-ray sources coincident with submm sources surrounding three HzRGs. 

These proto-clusters also contain a large amount of gas, as revealed by the \Lya\ haloes surrounding the HzRGs, which have physical scales up to $>$200~kpc \citep[for a recent review, see][]{wvb03}. Together with the increased merger rates in higher redshift clusters \citep[\eg][]{dok99}, this provides the ingredients to induce wide-spread starbursts and AGN, which could be revealed by their thermal dust (sub-)mm emission. Statistical overdensities of (sub-)mm galaxies (SMGs) have indeed been found from SCUBA bolometer imaging of the fields surrounding the $z$=3.09 'redshift spike' \citep{cha01}, the $z$=2.39 radio galaxy 53W002 \citep{sma03a}, the $z$=3.8 radio galaxy 4C~41.17 \citep{ivi00}, a $z$=1.8 QSO \citep{ste04} and six other HzRG fields \citep{ste03}.
In this paper, we present a 1.2~mm map covering the central 25~arcmin$^2$ of the most distant proto-cluster known to date, surrounding the $z$=4.1 radio galaxy TN~J1338$-$1942. We find an overdensity of 1.2~mm sources, which we identify with optically very faint galaxies using a deep 1.4~GHz map, but find no overlap between the population of excess \Lya\ emitters and the 1.2~mm sources.
Throughout this paper, we use a $\Lambda-$cosmology with H$_0$=71~km~s$^{-1}$~Mpc$^{-1}$, $\Omega_{\rm M}$=0.27 and $\Omega_{\Lambda}$=0.73 \citep[][]{spe03,ton03}. At $z$=4.1, the luminosity distance is $D_L$=37.65~Gpc, and 1\arcsec\ corresponds to 7.0~kpc. 

\section{Observations and data reduction}

\subsection{IRAM 1.2mm imaging}
To image the field of TN~J1338$-$1942 at mm wavelengths, we used the 37- and 117-channel Max Planck Bolometer arrays \citep[MAMBO-1 and MAMBO-2;][]{kre98} at the IRAM 30m telescope on Pico Veleta, Spain. MAMBO has a half-power spectral bandwidth from 210 to 290~GHz, with an effective bandwidth centre for steep thermal spectra of $\sim$250~GHz (1.2~mm). The effective beam FWHM is 10\farcs7 with an array size of 4\arcmin.

The observations were done in a pooled observing mode during the winter 2001-2002 season. Due to the low declination, the field could only be observed for 4 hours in a given night, with elevations between 29\degr\ and 33\degr. The atmospheric zenith opacities at 1.2~mm varied between 0.12 and 0.25. The total on source integration time was 17.0 hours, of which 2.6~h were obtained with the 117-channel array, and 14.6~h with the 37-channel array. 
We used the standard on-the-fly mapping technique, comprised of 41 subscans of 40~s each, while chopping the secondary mirror in azimuth at 2~Hz.
To minimize the residual effects of the double beam point spread function, we used different chop (wobbler) throws (39\arcsec, 42\arcsec\ or 45\arcsec) and/or scan directions for each map. 
We used 7 different pointings offset by 80\arcsec\ to cover the entire field of the VLT/FORS2 imaging to uniform depth. We checked the pointing and focus at least once per hour using the bright point source 1334$-$127, and found the pointing to be stable to within $<$2\arcsec. The absolute flux calibration is based on observations of several standard calibration sources, including planets, resulting in an estimated accuracy of 15\%. 
 
\begin{figure*}[ht]
\psfig{file=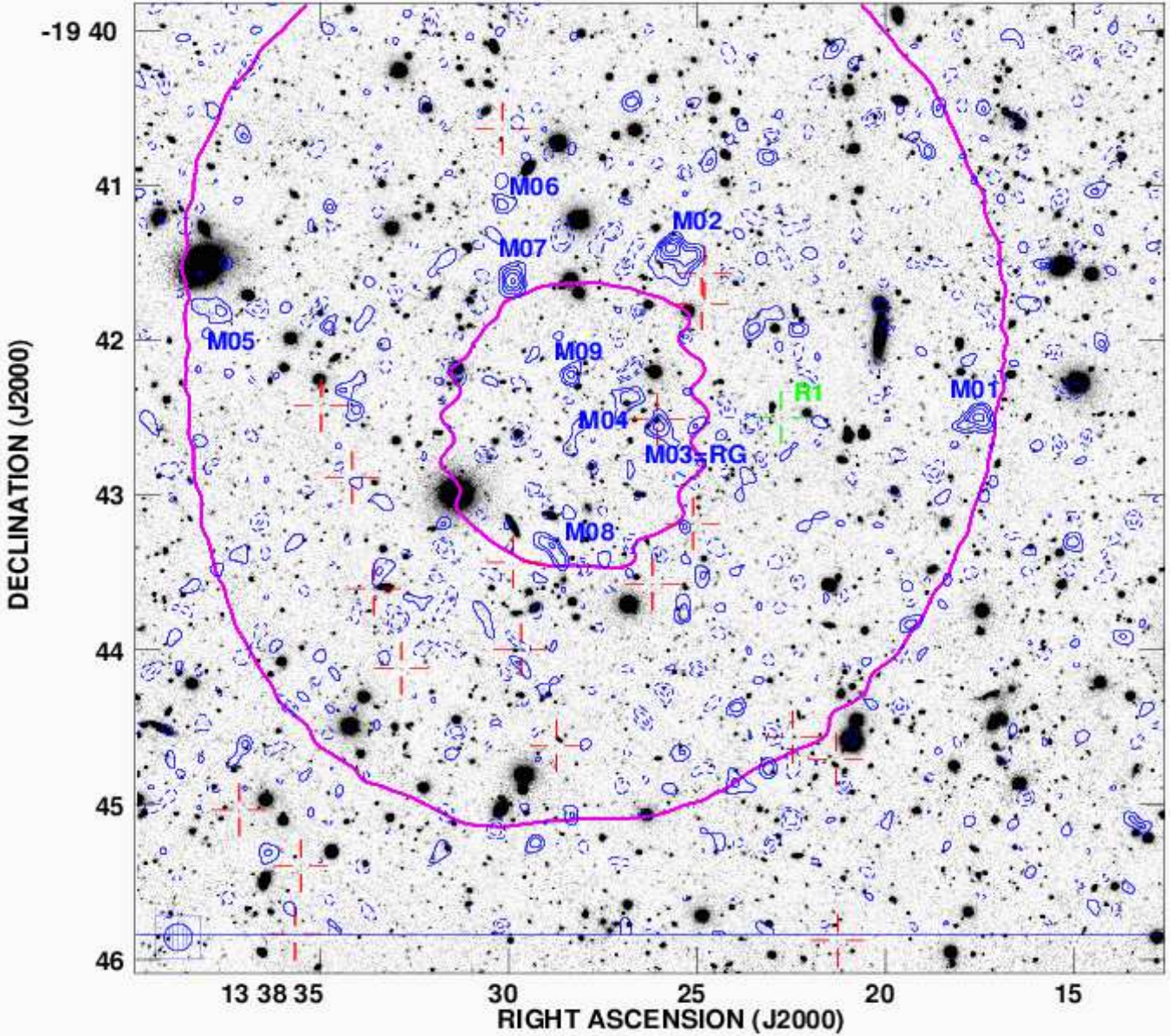,width=18cm}
\caption{VLT $R-$band image \citep[greyscale;][]{ven02} with MAMBO 1.2~mm signal-to-noise map (smoothed to 11\arcsec) overlaid as thin/blue contours. Contour levels are -3, -2, 2, 3, 4 and 5$\sigma$, with $\sigma$ the local rms noise level (negative contours are dashed). The two large purple contours delineate the regions with $\sigma<$0.6 and 1.2~mJy/beam. The MAMBO beam size is indicated in the lower left corner. The red open crosses indicate spectroscopically confirmed \Lya\ companions to the radio galaxy TN~J1338$-$1942, and the green open cross labeled R1 an additional FR~II radio source.}
\label{RMAMBOSN}
\end{figure*}

We analyzed the data using the MOPSI software package \citep{zyl98}. We subtracted the skynoise, and combined the double-beam maps using a shift-and-add procedure, producing for each map a positive image bracketed by two negative images of half the intensity located one chop throw away. Because we combined our 37 maps obtained with different chop throws, the effect of the confusion due to negative sidelobes is minimized, but still present in regions of high source density. The noise level increases outward in our co-added map of the field. In Fig.~\ref{RMAMBOSN} two contours show the region within which the rms noise level is less than 1.2 and 0.6 mJy (before smoothing), enclosing areas of 25.6 and 2.6 square arcmin, respectively.

\subsection{VLA radio imaging}
To obtain accurate positions of the mm sources, and to search for possible radio-loud AGN counterparts, we observed the TN~J1338$-$1942 field with the Very Large Array \citep[VLA; ][]{nap83} on UT 2002 April 1 to 12 for a total of 12~hours in the A-array at 20~cm. We observed in a pseudo-continuum, spectral line mode with $7 \times 3.125$~MHz channels. We monitored the point source 1351$-$148 every 40~min to provide amplitude, phase and bandpass calibration, and used an observation of 3C~286 to provide the absolute flux calibration.

We performed standard spectral-line calibration and editing of the data using the NRAO \aips\ package, and employed standard wide field imaging techniques \citep{tay99}. The final 7\farcm5$\times$7\farcm5 image has an rms noise level of 15~$\mu$Jy~beam$^{-1}$, except in the area close to the central radio galaxy, which is limited by the ability to clean the bright radio source (see Fig.~\ref{KMAMBOSNVLA}). The FWHM resolution of the restoring beam is $2\farcs3 \times 1\farcs3$ at a position angle PA=0\degr. 

\begin{table*}
\caption{Astrometry of the MAMBO, VLA and VLT sources in the TN~J1338$-$1942 field.}
\label{MAMBOastrometry}
\begin{tabular}{rllllll}
\hline
 & \multicolumn{2}{c}{MAMBO} & \multicolumn{2}{c}{VLA} & \multicolumn{2}{c}{VLT} \\
Source & RA(J2000) & DEC(J2000) & RA(J2000) & DEC(J2000) & RA(J2000) & DEC(J2000) \\
 & $\;\;^h\;\;\; ^m\;\;\; ^s\;\;\,$ & $\;\;\;\;\;$\degr$\;\;$ \arcmin$\;\;\;$ \arcsec$\;$ & $\;\;^h\;\;\; ^m\;\;\; ^s\;\;\,$ & $\;\;\;\;\;$\degr$\;\;$ \arcmin$\;\;\;$ \arcsec$\;$ & $\;\;^h\;\;\; ^m\;\;\; ^s\;\;\,$ & $\;\;\;\;\;$\degr$\;\;$ \arcmin$\;\;\;$ \arcsec$\;$ \\ 
\hline
M01 &13 38 17.56 &$-$19 42 30.7 & 13 38 17.48 &$-$19 42 29.5 & 13 38 17.43 &$-$19 42 29.9 \\
M02a&13 38 25.65 &$-$19 41 25.5 & 13 38 25.47 &$-$19 41 21.5 & 13 38 25.45 &$-$19 41 22.1 \\
M02b&            &              & 13 38 25.37 &$-$19 41 26.5 &             &              \\
M02c&            &              & 13 38 25.81 &$-$19 41 26.2 &             &              \\
M03 &13 38 25.97 &$-$19 42 33.9 & 13 38 26.10 &$-$19 42 31.3 & 13 38 26.06 &$-$19 42 30.7 \\
M04 &13 38 26.84 &$-$19 42 22.8 & 13 38 26.83 &$-$19 42 25.8 & 13 38 26.82 &$-$19 42 26.2 \\
M05 &13 38 37.79 &$-$19 41 48.8 & 13 38 37.91 &$-$19 41 49.1 & 13 38 37.92 &$-$19 41 48.9 \\
M06 &13 38 30.13 &$-$19 41 06.5 & 13 38 30.04 &$-$19 41 04.5 & 13 38 30.06 &$-$19 41 04.7 \\
M07 &13 38 29.89 &$-$19 41 36.9 & \nodata     &\nodata       & \nodata     &\nodata       \\
M08 &13 38 28.86 &$-$19 43 20.9 & \nodata     &\nodata       & 13 38 28.77 &$-$19 43 27.5 \\
M09 &13 38 28.33 &$-$19 42 13.2 & \nodata     &\nodata       & \nodata     &\nodata       \\
M10 &13 38 29.78 &$-$19 39 32.1 & 13 38 29.81 &$-$19 39 31.9 & \nodata$^a$ & \nodata$^a$  \\
R1  &\nodata     &\nodata       & 13 38 22.77 &$-$19 42 30.0 & 13 38 22.76 &$-$19 42 29.9 \\
\hline
\end{tabular}

$^a$ This source is located outside the VLT image boundary.
\end{table*}

\begin{table*}
\caption{Photometry of the MAMBO and VLA sources in the TN~J1338$-$1942 field.}
\label{MAMBOphotometry}
\begin{tabular}{rrrrrccc}
\hline
Source & $S_{\rm 20cm}$  & $S_{\rm 1200\mu m}$ & $S_{\rm 850\mu m}$ & $S_{\rm 450\mu m}$ & $R$[3\arcsec] & $NB$[3\arcsec] & $K$[3\arcsec] \\
 & $\mu$Jy & mJy/beam & mJy & mJy & mag & mag & mag \\ 
\hline
M01 &{\bf 148$\pm$34}    &{\bf 6.2$\pm$1.2}  &{\bf 10.1$\pm$1.3}    &{\bf 21.5$\pm$6.4}& 26.8$\pm$0.3  & $>$28        &\nodata$^a$\\
M02a&      54$\pm$25     &{\bf 4.1$\pm$0.8}  &{\bf  6.6$\pm$1.3}$^b$&  12.5$\pm$6.0$^b$& 25.6$\pm$0.1 & 26.7$\pm$0.7& \nodata$^a$\\
M02b&      46$\pm$25     &                   &                      &                  & $>$28         & $>$28        &\nodata$^a$\\
M02c&      47$\pm$26     &                   &                      &                  & $>$28         & $>$28        &\nodata$^a$\\
M03&{\bf 120800$\pm$4300}&{\bf 2.3$\pm$0.6}  &{\bf  6.1$\pm$1.3}$^c$&     25.3$\pm$9.3 & 22.38$\pm$0.01& 19.2$\pm$0.1 &19.97$\pm$0.05\\
M04 &      47$\pm$17     &{\bf 2.0$\pm$0.5}  & \nodata              & \nodata          & 24.08$\pm$0.03& 24.9$\pm$0.2 &18.72$\pm$0.03\\
M05 &{\bf  91$\pm$25}    &{\bf 3.8$\pm$0.8}  &{\bf  9.9$\pm$1.2}$^b$&  14.9$\pm$5.9$^b$& 25.06$\pm$0.07& 25.6$\pm$0.3 &\nodata$^a$\\
M06 &      41$\pm$24     &{\bf 2.3$\pm$0.6}  &{\bf  5.7$\pm$1.3}    &$<$20 (3$\sigma$) & 27.8$\pm$0.8  & $>$28        &\nodata$^a$\\
M07 &$<$50 (3$\sigma$)   &{\bf 4.0$\pm$0.6}  &      3.3$\pm$2.0     &$<$33 (3$\sigma$) & $>$28         & $>$28        &\nodata$^a$\\
M08 &$<$50 (3$\sigma$)   &{\bf 2.4$\pm$0.7}  & \nodata              & \nodata          & 25.26$\pm$0.09& 26.0$\pm$0.4 &19.2$\pm$0.1\\
M09 &$<$50 (3$\sigma$)   &{\bf 2.3$\pm$0.6}  & \nodata              & \nodata          & \nodata$^d$   & \nodata$^d$  &\nodata$^a$\\
M10 &{\bf 195$\pm$28}    &{\bf 3.2$\pm$1.0}  & \nodata              & \nodata          & \nodata$^a$   & \nodata$^a$  &\nodata$^a$\\
R1  &10900$\pm$400       &$<$1.8 (3$\sigma$) & \nodata              & \nodata          & 25.6$\pm$0.1  & 25.6$\pm$0.3 &\nodata$^a$\\
\hline
\end{tabular}

$^a$ Source located outside the coverage of the VLT/Keck images.

$^b$ SCUBA flux densities could be underestimated due to the extent of the 1.2mm source (see \S3.2 and Fig.~\ref{RMAMBOSNVLAIDs}).

$^c$ We quote the S/N weighted average of our value ($S_{\rm 850\mu m}=3.1\pm 1.8$) and the value quoted by \citet{reu04} ($S_{\rm 850\mu m}=6.9\pm 1.1$).

$^d$ No clear optical/near$-$IR counterpart.
\end{table*}

\subsection{JCMT submm photometry}
We obtained 850~$\mu$m and 450~$\mu$m photometry of 6 MAMBO sources previously identified using the VLA and VLT imaging (\S 3.1) using the Submillimetre Common-User Bolometer Array \citep[SCUBA;][]{hol99} on the 15m James Clerk Maxwell Telescope on UT 2003 February 17 to 22, for a total integration time of 16~hours (including overheads) spread over 6 sources. The respective beamsizes are 14\farcs7 at 850~$\mu$m and 7\farcs5 at 450~$\mu$m. We used the recommended 9-point jiggle photometry mode, which samples a $3 \times 3$ grid with 2\arcsec\ spacing between grid points, while chopping in azimuth by 60\arcsec.
The atmospheric conditions were excellent, with zenith opacities at 850~$\mu$m varying between 0.11 and 0.24. We checked the pointing and focus roughly once per hour using the bright point source 1334$-$127, and found the pointing to be stable within $<$3\arcsec. The absolute flux calibration is based on observations of CRL618 and Mars.

We reduced the data in the standard way using SCUBA User Reduction Facility \citep[SURF;][]{jen97}. We removed individual data samples that were deviant from the mean by $>$4$\sigma$ from the data set, flatfielded, corrected for extinction, despiked, and subtracted the residual sky level from the off-source bolometers. Finally, we calculated the mean and rms from the time sequence to remove all points deviating by more than 3$\sigma$ from the mean. We adopt the mean of the remaining data as our final estimate of the intensity. 

\subsection{Optical and near$-$IR imaging and spectroscopy}
Optical and near-IR imaging data of the TN~J1338$-$1942 field exist in the literature. \citet{deb99} published a NIRC/Keck $K-$band image covering the central 1\arcmin$\times$1\arcmin, centred on the radio galaxy. \citet{ven02} published deep $R-$band and narrow-band \Lya\ images obtained with the FOcal Reducer/low dispersion Spectrograph (FORS) at the ESO Very Large Telescope (VLT), covering 6\farcm4$\times$6\farcm2, and also spectroscopically confirmed 20 \Lya\ emitters in this field, of which 15 fall within the $\sigma<1.2$~mJy region of our MAMBO map. Overzier~\etal~(in prep.) also obtained a $K-$band image of the south-eastern corner of the proto-cluster with the Infrared Spectrometer And Array Camera (ISAAC) at the VLT.  In the following, we shall use both the $K-$band and $R-$band images to identify the optical counterparts of the mm and radio sources. We corrected all magnitudes for Galactic extinction $E(B-V)$=0.097 using the dust maps of \citet{schl98} and the extinction curve of \cite{car89}.

Optical spectroscopy of three sources (M01, M05 and M08) was obtained as part of two VLT multi-slit spectroscopy programs. Details about these observations will be given in future papers (De Breuck \etal, in preparation; Overzier \etal, in preparation).

\begin{figure*}[ht]
\psfig{file=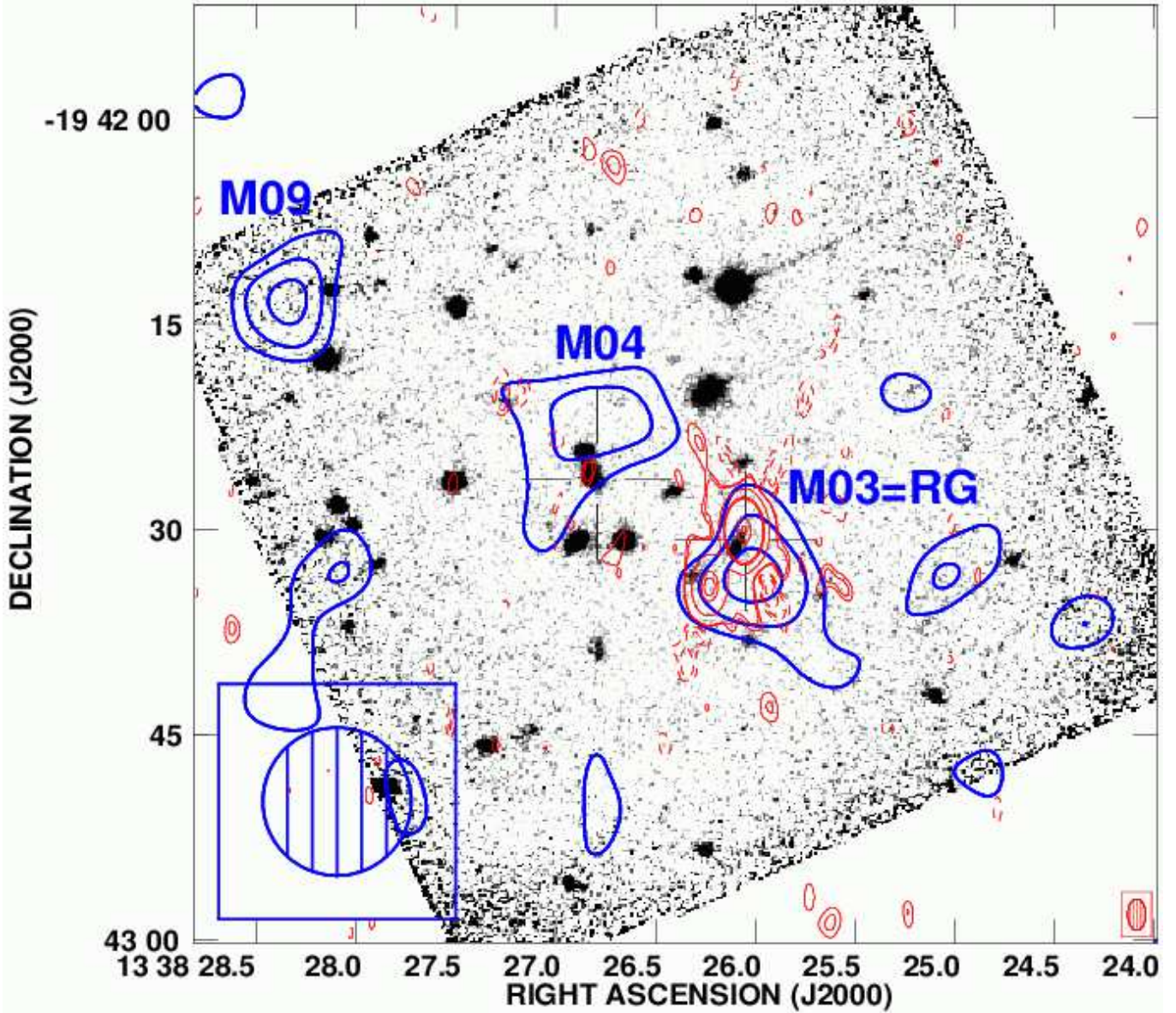,width=18cm}
\caption{Keck/NIRC $K-$band image (greyscales) with the MAMBO 1.2~mm signal-to-noise map (smoothed to 11\arcsec) overlaid as thick/blue contours and the VLA 1.4~GHz map as thin/red contours. Contour levels for the MAMBO map are 2, 3 and 4$\sigma$, with $\sigma$ the local rms noise level. Contour levels for the VLA map are -0.1275, -0.09, -0.063, -0.045, 0.045, 0.063, 0.72, 4.08, 5.775, 62.28 and 130.56 mJy/beam. The MAMBO and VLA beam sizes are indicated in the lower left and lower right corners, respectively. The open cross indicates the position of the optical/radio identification.}
\label{KMAMBOSNVLA}
\end{figure*}

\section{Results}

\subsection{Extraction of the mm sources}
We use the MAMBO signal-to-noise map (Fig.~\ref{RMAMBOSN}) to identify candidate mm sources within the region where $\sigma <$1.2~mJy/beam. To assess the reliability of the candidate sources, we also made 'probability maps' by splitting up the data sets, and looking for gross discrepancies between them. This would identify spurious sources appearing in only part of the data. The reliability of our positive detections is further strengthened by the absence of negative peaks $<-4\sigma$ in the S/N map (dashed contours in Fig.~\ref{RMAMBOSN}). The three most significant negative sources are all located in between M02 and M07, consistent with them being residual double beaming effects (see \S 2.1).

Table~\ref{MAMBOastrometry} lists the positions of 10 candidate 1.2~mm sources found in the MAMBO map, and Table~\ref{MAMBOphotometry} their flux densities. We fitted the sources in the map smoothed to 11\arcsec\ to Gaussians unconstrained in width, and in cases where the fitted width is smaller than the beam size, to constrained width Gaussians of 11\arcsec\ FWHM. The peak fluxes quoted in Table~\ref{MAMBOphotometry} are averages of Gauss fits in maps smoothed to 11\arcsec\ and 12\arcsec, and to eliminate a base problem due to double beams surrounding the brighter sources, in a map truncated toward negative values at 0. The quoted uncertainties are quadratic sums of the dispersion in these three fits and the rms noise level at the position of the source in the 11\arcsec\ resolution map.

\begin{figure*}[ht]
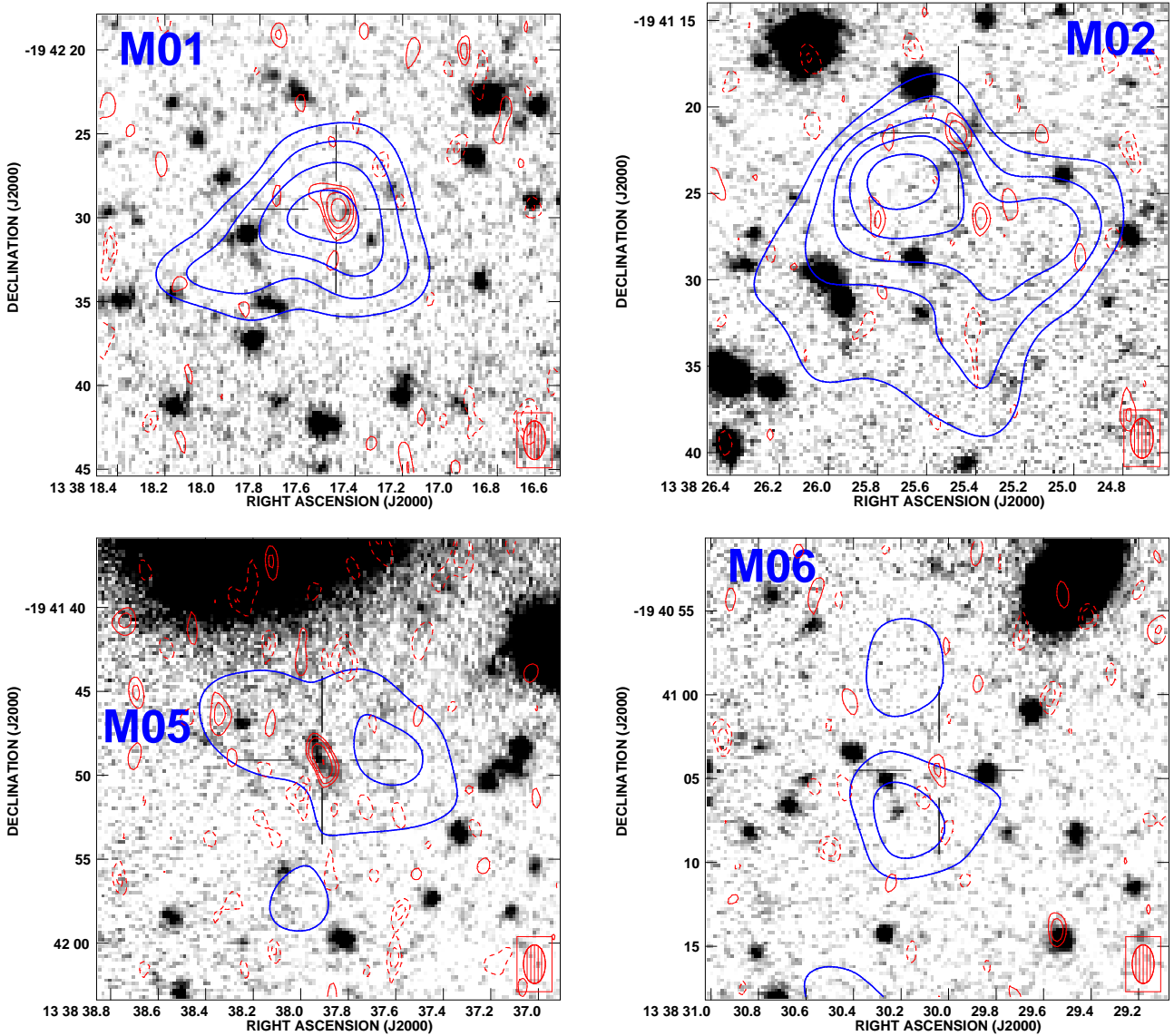

\begin{tabular}{ll}
\psfig{file=debreuck.fig3a.ps,width=8.5cm,angle=-90} &
\psfig{file=debreuck.fig3b.ps,width=8.5cm,angle=-90} \\
\psfig{file=debreuck.fig3c.ps,width=8.5cm,angle=-90} &
\psfig{file=debreuck.fig3d.ps,width=8.5cm,angle=-90} \\
\end{tabular}
\caption{VLT $R-$band image (greyscales) with the MAMBO 1.2~mm signal-to-noise map (smoothed to 11\arcsec) overlaid as thick/blue contours and the VLA 1.4~GHz map as thin/red contours. Contour levels for the MAMBO map are 2, 3, 4, 5, 6 and 7$\sigma$, with $\sigma$ the local rms noise level. Contour levels for the VLA map are -42,-30, 30, 42, 60 and 85~$\mu$Jy/beam. The VLA beam size is indicated in the lower right corner. The open cross indicates the position of the optical/radio identification.}
\label{RMAMBOSNVLAIDs}
\end{figure*}
\addtocounter{figure}{-1}
\begin{figure*}[ht]
\begin{tabular}{ll}
\psfig{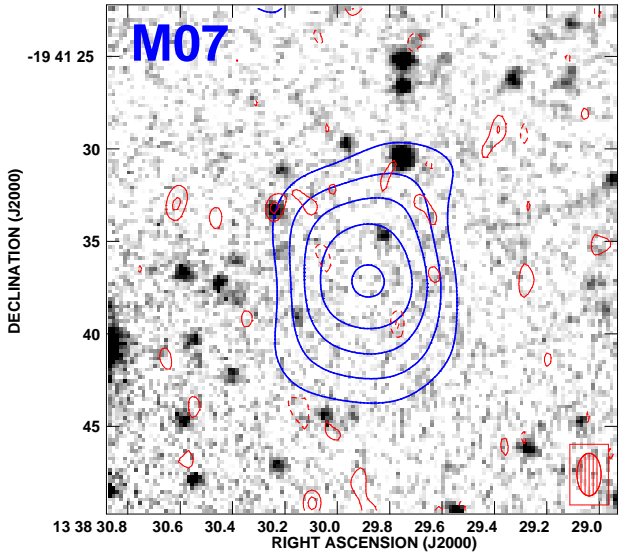} &
\psfig{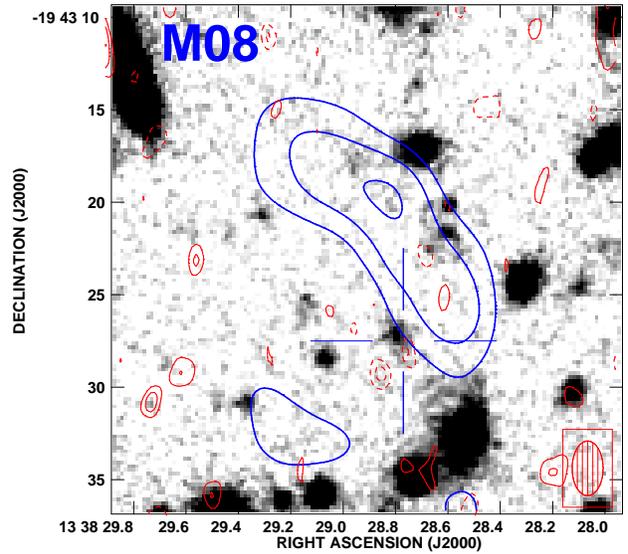} \\
\end{tabular}
\caption{, {\it continued}}
\end{figure*}

\subsection{Identification of the mm sources}
Due to the large MAMBO beam (FWHM $\sim 11$\arcsec), we often find several potential optical counterparts within the positional uncertainty. Higher resolution interferometric imaging would be needed for more unique identifications \citep[\eg][]{dan02,dun03}. Comparisons of positions derived from MAMBO maps and from the VLA or IRAM Plateau de Bure interferometer yield median offsets of 2--3\arcsec\ with a tail extending to 7--8\arcsec\ \citep{eal03,dan04}. 
To obtain more accurate positions, we can take advantage of the tight radio-to-far-infrared correlation in star-forming galaxies \citep[\eg][]{con92}, which appears to hold out to high redshift \citep{gar02}. Because the faint radio synchrotron emission traces the same region as the mm emission, we can use deep sub-arcsecond resolution 1.4~GHz VLA maps to pinpoint the dust emission to within $<$1\arcsec\ \citep[\eg][]{ivi02}.

The surface density of radio sources brighter than 30~$\mu$Jy is $\simeq$2.5~arcmin$^{-2}$ \citep{ivi02}, so we expect to find only 0.035 such sources within a radius of 4\arcsec\ from the MAMBO positions. Because 30$\mu$Jy is at the 2$\sigma$ noise level in our radio map, we expect to find 0.4 positive 2$\sigma$ peaks within 4\arcsec\ from the nominal MAMBO positions. In an attempt to discriminate against noise peaks, we also checked for optical or near--IR counterparts for the radio sources. We consider an optical/near--IR source with a $>$2$\sigma$ radio peak within the astrometric uncertainty of 0\farcs4 as a likely counterpart to a MAMBO source, if it lies within 4\arcsec\ from the MAMBO position. In the following, we discuss the individual identifications of the 10 candidate MAMBO sources, and of an additional radio source in the field. Table~\ref{MAMBOastrometry} lists the mm, radio and optical positions, and Table~\ref{MAMBOphotometry} the photometric measurements.

\noindent{\bf M01} (Fig.~\ref{RMAMBOSNVLAIDs}): This is the brightest source in our MAMBO map. The identification is secure with a 4.4$\sigma$ radio source coinciding with an $R$=26.8 source, which we used as the basis for SCUBA photometry. We have attempted deep VLT/FORS1 spectroscopy on this object (De Breuck \etal\ in preparation), but did not detect any emission, so the spectrum is not dominated by strong line emission from an AGN. 

\noindent{\bf M02} (Fig.~\ref{RMAMBOSNVLAIDs}): This MAMBO source appears to consist of two components; an unconstrained Gaussian fit yields a size of 19\farcs7$\times$17\farcs0. Within a radius of 5\arcsec\ from the MAMBO position, there are three 1.4~GHz sources with flux densities $S_{\rm 1.4GHz}>$45~$\mu$Jy, i.e. at significance levels of $\sim$2$\sigma$ (see Table~\ref{MAMBOphotometry}), while we expect to find only one such source. This suggests a complex, merging structure of galaxies, as hinted by the diffuse $R-$band source coincident with the brightest of the three VLA sources, M02a (see Tables~\ref{MAMBOastrometry} and \ref{MAMBOphotometry}). We have used this position for the SCUBA photometry, but the relatively low $S_{850}/S_{1200}$ ratio indicates that the SCUBA beam may have missed part of the flux, suggesting M02a is not the correct identification of the brightest 1.2~mm emission. Only interferometric (sub-)mm imaging could provide a better understanding of this interesting source.

\noindent{\bf M03} (Fig.~\ref{KMAMBOSNVLA}): This 3.8$\sigma$ MAMBO source lies 3\farcs2 from the host galaxy of the powerful radio source \citep{deb99}, and is therefore most likely related. We used the $K-$band position to obtain MAMBO on-off observations and SCUBA photometry, but both of these gave lower than expected flux densities (see Table~\ref{MAMBOphotometry}), suggesting that the dust emission could be offset from the $K-$band position, as has been seen in the $z$=3.79 radio galaxy 4C~60.07 \citep{pap00}. In Table~\ref{MAMBOphotometry}, we quote the S/N weighted average of our $S_{\rm 850 \mu m}$=3.1$\pm 1.8$~mJy measurement and the $S_{\rm 850 \mu m}$=6.9$\pm 1.1$~mJy of \citet{reu04}. We do not average the 450~$\mu$m flux densities, as the \citet{reu04} photometry was obtained under adverse atmospheric conditions leading to a nominal measurement of $S_{\rm 450 \mu m}=-36.3\pm 31.5$~mJy. Such discrepant $S_{\rm 850 \mu m}$ flux densities between observations at different epochs have been noticed before for MG~2141+192 \citep{arc01,reu04}, and suggest submm variability. But, as \citet{reu04} argue, the extended nature of the submm emission excludes this explanation, unless there is a significant contribution from non-thermal AGN emission. However, a power-law extrapolation from the total 8.2~GHz flux density \citep{pen00b} predicts a negligible AGN synchrotron contribution of only 30~$\mu$Jy at 250~GHz, but given the unusually asymmetric morphology of the radio source \citep{deb99}, we cannot exclude the presence of a variable, flat spectrum radio core contributing to the submm flux densities, as has been seen in B2~0902+343 \citep{dow96}. High resolution imaging at frequencies of several tens of GHz would be needed to test this.

\noindent{\bf M04} (Fig.~\ref{KMAMBOSNVLA}): This 4$\sigma$ MAMBO source lies 3\farcs0 from a faint radio source coincident with a very red galaxy with $R-K$=5.36$\pm$0.04, which is a likely identification.

\noindent{\bf M05} (Fig.~\ref{RMAMBOSNVLAIDs}): We obtained SCUBA photometry of the extended radio and optical source, confirming the identification. This is the only source which has spatially extended 1.4~GHz emission; the close correspondence between the diffuse optical and radio morphologies strongly suggests the 1.4~GHz emission traces starburst rather than AGN emission. Deep VLT/FORS1 spectroscopy (De Breuck \etal\ in preparation) also revealed a faint featureless continuum, but no emission lines. An unconstrained Gaussian fit to the MAMBO S/N map yields a size of 20\farcs8$\times$18\farcs0, but the S/N of our detection is only 3.7 in a 11\arcsec\ smoothed map, so the detection of spatially extended emission is tentative at best.

\noindent{\bf M06} (Fig.~\ref{RMAMBOSNVLAIDs}): We obtained SCUBA photometry at the position of a very faint radio and optical identification, confirming the reality of this faint MAMBO source. Note that the apparent spatial extent of the MAMBO emission cannot be trusted because this source is only detected at the 3.4$\sigma$ level.

\noindent{\bf M07} (Fig.~\ref{RMAMBOSNVLAIDs}): Although this is one of the brightest MAMBO sources, there is no obvious radio or optical identification. We have obtained SCUBA photometry of the radio/optical source at RA=13$^h$38$^m$30.22$^s$, DEC=$-$19\degr41\arcmin33\farcs06, but obtained only a 1.5$\sigma$ signal of $S_{\rm 850\mu m}$=3.7$\pm$2.5~mJy, while at the nominal MAMBO position, we obtain a slightly higher signal (see Table~\ref{MAMBOphotometry}). We do not have a good candidate optical or radio counterpart for this MAMBO source.

\noindent{\bf M08} (Fig.~\ref{RMAMBOSNVLAIDs}): This faint MAMBO source lies 6\farcs7 from an extremely red object (ERO) with $K$=19.2$\pm$0.1 and $R-K$=6.1$\pm$0.15. We have obtained deep FORS2/MXU spectroscopy of this ERO (Overzier \etal, in preparation), detecting a faint red continuum and an emission line at 8120\AA, which we tentatively identify as \OII\ at $z$=1.18. Note that from blank-field ERO surveys, the density of objects with $R-K$$>$6 and $K$$<$19.2 is 0.10 arcmin$^{-2}$ \citep{dad00}, so the chance of finding such an object within 6\farcs7 is $P$=0.4\%. It is therefore possible that M08 is a dusty ERO \citep[\eg][]{cim98,dey99,sma02b,tak03}. However, because the ERO is not the closest possible identification, it is statistically not the most likely identification.  Note that the apparent spatial extent of the MAMBO emission cannot be trusted because this source is only detected at the 4$\sigma$ level. This may also indicate that this source has made it into our sample due to the confusion of two sources too close to be detected individually. Deeper mm/submm observations would be needed to verify this, and to determine if the dust emission is related to the ERO or not.

\noindent{\bf M09} (Fig.~\ref{KMAMBOSNVLA}): We find no radio source within one MAMBO beamsize, and no clear optical/near$-$IR identifications.

\begin{figure}[ht]
\psfig{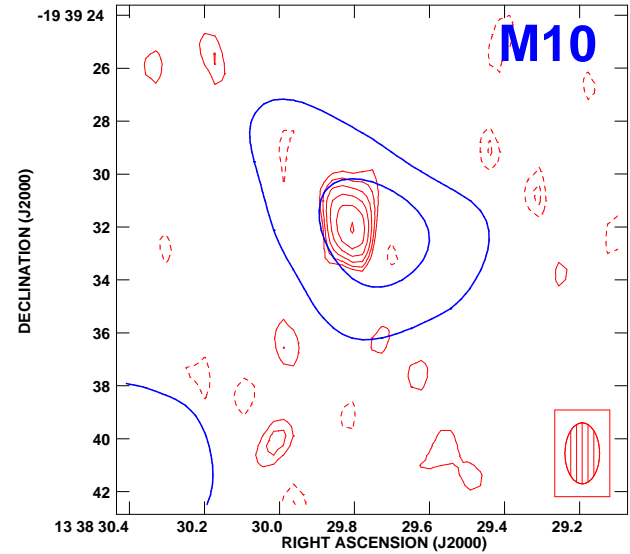}
\caption{M10: MAMBO 1.2~mm signal-to-noise map (thick/blue contours; smoothed to 11\arcsec) and the VLA 1.4~GHz map (thin/red contours). Contour levels for the MAMBO map are 2 and 3$\sigma$, with $\sigma$ the local rms noise level. Contour levels for the VLA map are -42,-30, 30, 42, 60, 85, 120 and 170~$\mu$Jy/beam. The VLA beam size is indicated in the lower right corner. This source falls outside the VLT $R-$band image.}
\label{RMAMBOSNVLAnoIDs}
\end{figure}

\noindent{\bf M10} (Fig.~\ref{RMAMBOSNVLAnoIDs}): This MAMBO source falls just outside the VLT $R-$band image. It coincides with a strong 0.2~mJy radio source.

\noindent{\bf R1} (Fig.~\ref{RMAMBOSN}): This is a moderately bright radio source with a FR~II \citep{fan74} morphology. We detect no 1.2~mm emission from this source.

\subsection{Extended 1.2mm emission}

At least one (M02) of the three sources detected with S/N$>$5, and possibly the weaker source M05 appear to have significantly extended 1.2~mm emission. M02 appears to consist of two barely resolved components, though none of them have a clear candidate optical or radio identification. Spatially extended submm emission has been reported before in several HzRGs and at least two companion sources \citep[][Ivison \etal, in prep.]{ivi00,ste03}. This suggests that star formation in these objects occurs over scales of several tens of kpc.

A possible alternative explanation is that M02 (and maybe M05) could consist of multiple images by strong gravitational lensing by a foreground cluster \citep{kne04,bor04}. However, we have not detected any gravitational arcs in the ACS/{\it HST} images \citep{mil04} or in the VLT image (Fig.~\ref{RMAMBOSN}). We therefore consider it extremely unlikely that gravitational lensing can explain the spatial extent in the mm emission. A more detailed analysis of the lensing properties, including an analysis of a possible weak shear using the ACS image will be presented in a future paper (Overzier \etal, in prep.).

\subsection{Surface density}

Although many of the MAMBO sources were detected at low S/N level, most of them appear to be real because because they have been confirmed with pointed sub-mm photometry, or because they have plausible counterparts at radio and/or optical wavelengths. We detect 10 sources brighter than 2.0~mJy (peak S/N$>$3, a level at which we should be close to being complete). Omitting M03, which corresponds to the pre-selected radio galaxy in the field, this corresponds to a surface density of 0.35~arcmin$^{-2}$ (roughly 1\% of the confusion limit). Note that this value is likely to be an overestimate because we consider sources down to the 3$\sigma$ level, which may have lead to flux boosting \citep[\eg][]{eal03}. This effect raises intrinsically fainter sources above the detection thres\-hold due to the addition of noise (instrumental and atmospherical) or due to confusion of sources too faint to be detected individually. The latter affect may be happening in M08 (Fig.~\ref{RMAMBOSNVLAIDs}). \citet{eal03} argue that for MAMBO surveys, the average boosting factor is 14\%, which is slightly lower than for the SCUBA surveys, possibly due to the smaller MAMBO beam.

We now compare our surface density with the integrated 1.2~mm source counts from the MAMBO blank field surveys \citep{ber00}.
Using the 1.2~mm counts from Bertoldi (private communication) and \citet{gre04}, our $S_{\rm 1.2mm}$ source density (without flux boosting correction) is roughly twice as high as expected from the random-field density. However, if we consider only the 3 sources with $S_{\rm 1.2 mm}>$4~mJy, the source density is 0.12~arcmin$^{-2}$, three times higher than expected. Assuming a Poissonian distribution, this represents a 7$\sigma$ overdensity. These bright sources are all detected with a local\footnote{Note that the uncertainties in Table~\ref{MAMBOphotometry} also include the fitting uncertainties (see \S3.1), and are hence larger than the local rms.} S/N$>$5, so they will not be affected by flux boosting. 
This overdensity suggests that most of the brighter MAMBO sources and a few of the fainter ones may be related to the proto-cluster. Such statistical overdensities of SMGs are also seen in SCUBA maps of other HzRG fields \citep[\eg][]{ste03}. In the next section, we use our radio, mm and submm photometry to constrain the redshifts of these MAMBO sources.
 
\begin{table}
\caption{Photometric redshift estimates.}
\label{zestimates}
\begin{tabular}{ccc}
\hline
Source & $z_{\rm est}$ from $\alpha_{\rm 1.4 GHz}^{\rm 250 GHz}$ & $z_{\rm est}$ from $S_{\rm 850\mu m}/S_{\rm 1200\mu m}$ \\
\hline
M01     & 2.2$^{+1.7}_{-0.9}$ & 6.8$^{+\infty}_{-4.5}$\\
M02     & 2.8$^{+3.5}_{-1.3}$ & 7.1$^{+\infty}_{-4.9}$\\
M03$^a$ & \nodata             & 7.1$^{+11}_{-1.9}$    \\
M04     & 2.2$^{+2.1}_{-1.1}$ & \nodata               \\
M05     & 2.2$^{+1.8}_{-1.0}$ & 2.1$^{+6.8}_{-2.1}$   \\
M06     & 2.2$^{+4.1}_{-1.0}$ & 2.4$^{+14}_{-2.4}$    \\
M07     & $>$2.9              & $>$4.2                \\
M08$^b$ & $>$2.3              & \nodata               \\
M09     & $>$2.3              & \nodata               \\
M10     & 1.5$^{+1.1}_{-0.7}$ & \nodata               \\
\hline
\end{tabular}

$^a$ Radio galaxy at $z$=4.1.

$^b$ Possibly identified with a $z_{\rm spec}$=1.18 ERO, see \S3.2.
\end{table}

\subsection{Photometric redshift estimates}

We attempted very deep ($t_{\rm int}$=14h) optical spectroscopy of M01 and M05, but did not detect features to determine the redshifts. None of the MAMBO sources show excess narrow-band emission in the $4.075<z<4.123$ coverage of the \Lya\ filter, and only M04 and M08 have $K-$band photometry, so we cannot use conventional optical photometric redshift techniques. As this is a common situation for submm galaxies, several redshift estimators have been developed based on the submm and radio data only. 

\citet{car99} proposed to use the radio-to-submm spectral index as a redshift estimator, which is based on the local radio-to-far-infrared correlation. Table~\ref{zestimates} lists the redshift estimates calculated using the \citet{car00} $\alpha_{\rm 1.4 GHz}^{\rm 250 GHz}$ models. The quoted uncertainties include both the 1$\sigma$ measurement uncertainties in our photometry and the 1$\sigma$ scatter in the models. The estimated redshifts are clearly lower than the $z$=4.1 of the radio galaxy, though $z$=4.1 is still within the 1$\sigma$ uncertainties for most sources (except M10). However, a comparison with spectroscopic redshifts of SMGs indicates that the spectral-index redshifts tend to be systematically underestimated for $z_{\rm spec}>2$ objects \citep[\eg][]{cle04}. In fact, the \citet{car00} models predict a spectral index $\alpha_{\rm 1.4 GHz}^{\rm 250 GHz}$=1.00 for $z$=4.1, so for the average $S_{1200}$=3.3~mJy in our maps, we expect to find a radio source with a flux density $S_{\rm 1.4 GHz}$=18$\mu$Jy, well below the detection limit of our VLA map. However, \citet{pet03} report 1.4~GHz detections of $\sim$70$\mu$Jy in two $z>5$ radio quiet quasars with $S_{1200}$ flux densities of 0.9 and 5.5~mJy.
A possible explanation for this higher than expected 1.4~GHz emission is that an optically undetected AGN contributes to the radio emission. Note that AGNs have been spectroscopically confirmed in two proto-clusters surrounding HzRGs \citep{lef96,pen02}, while \citet{sma03b} report the discovery of {\it Chandra} X-ray sources coincident with submm sources surrounding three HzRGs, suggesting the MAMBO sources surrounding TN~J1338$-$1942 -- if they are part of the protocluster -- may well contain AGNs. Hence, a relatively bright radio detection of a MAMBO source does not exclude it as a potential member of the $z=4.1$ protocluster.

\citet{eal03} have also predicted the redshift evolution of the $S_{\rm 850\mu m}$/$S_{\rm 1200\mu m}$ ratio, by fitting a two-temperature model to a sample of 104 galaxies from the IRAS bright galaxy survey \citep{dun01}. We have used their median predicted value \citep[Fig.~4 in the ][paper]{eal03} to obtain an additional redshift estimate. Table~\ref{zestimates} lists these estimates; the quoted uncertainties include the 1$\sigma$ measurement uncertainties in the $S_{\rm 850\mu m}$/$S_{\rm 1200\mu m}$ ratios and the full spread in the model predictions. We find that, except for a $z>2.2$ limit for M01 and M02, this ratio does not provide a useful constraint on the redshifts due to (i) the small difference in wavelength between the mm and submm points and (ii) the relatively low S/N of our detections.

In summary, while the radio, mm and submm photometry are generally consistent with $z$=4.1, the uncertainties from these redshift estimate techniques are far too large to provide proof that the sources are at the redshift of the radio galaxy. Better sampled SEDs, especially in the submm would be needed to constrain the redshifts with photometry only.

\subsection{Statistical 1.2~mm and 1.4~GHz emission from the \Lya\ emitters and Mean Star Formation Rates}
None of the 14 spectroscopically confirmed \Lya\ emitters within the $\sigma<1.2$~mJy region (excluding the radio galaxy) is detected at $>2\sigma$ in our 1.2~mm map. To test if there is a statistical signal in our non-detections, we have stacked the emission from these 14 positions, and find $\langle S_{\rm 1.2mm}\rangle$=0.25$\pm$0.24~mJy. Using this 3$\sigma$ upper limit, and assuming dust parameters $T_{\rm d}$=33~K and $\beta$=2.0 used for LBGs \citep{bak01}, we derive $\langle L_{\rm FIR}\rangle \simlt 4.4 \times 10^{12}{\rm L_{\odot}}$, implying a mean star formation rate $\langle$SFR$_{\rm FIR}\rangle \simlt$~440~M$_{\odot}$yr$^{-1}$ \citep[\eg][]{omo01}. 

Similarly, from our VLA map, we find $\langle S_{\rm 1.4GHz}\rangle$=$-$0.2$\pm$2.9~$\mu$Jy. Using the relation between the radio emission and star formation rate \citep{con92}, we can use this 3$\sigma$ upper limit of 8.7~$\mu$Jy on the 1.4~GHz emission to calculate an upper limit to the star formation rate (SFR) from the \Lya\ emitters. Assuming a radio spectral index $\alpha_{\rm radio}$=$-$0.8, we derive $\langle$SFR$_{\rm radio}\rangle < $265~M$_{\odot}$yr$^{-1}$. 

The values should be compared with the SFR derived from the mean \Lya\ luminosity of the same 14 emitters (excluding the radio galaxy) $\langle L_{\rm Ly\alpha} \rangle$=2.77$\times$10$^{42}$erg~s$^{-1}$. Assuming a case~B ratio \Lya/\Ha=10, and using the \citet{ken94} relation between SFR and $L_{\rm H\alpha}$, we find $\langle$SFR$_{\rm Ly\alpha}\rangle$=2.2~M$_{\odot}$yr$^{-1}$. Note that this value is likely to be an underestimate, as \Lya\ is often quenched by dust emission, as illustrated by \citet{kurk04}, who report \Lya/\Ha\ ratios significantly lower than the case~B value for a sample of \Lya\ and \Ha\ emitters surrounding the $z$=2.16 radio galaxy PKS~1138$-$262. However, it is obvious that the deep \Lya\ imaging probes much lower SFR than the MAMBO and VLA maps.

\section{Discussion}
Although we could not put strong constraints on the redshifts of the 9 MAMBO sources surrounding TN~J1338$-$1942 using photometric redshift estimators or very deep VLT/FORS1 spectroscopy of two of them, the analysis of the source density and the photometric redshift estimates suggests half of the 9 MAMBO sources, and in particular the brightest ones, may well belong to the $z$=4.1 proto-cluster. To confirm (or refute) this requires alternative redshift determinations such as (i) deep near-IR spectroscopy, (ii) mid-IR spectroscopy with IRS/{\it Spitzer} using the PAH features, and (iii) mm spectroscopy using molecular CO lines.  

Figure~\ref{RMAMBOSN} shows that the nine MAMBO sources are not distributed uniformly around the radio galaxy, although the sensitivity of our MAMBO map radially decreases from the radio galaxy. The four brightest MAMBO sources (M01, M02, M07 and M05) are all north of the radio galaxy, while the densest area of \Lya\ emitters is located southeast of the radio galaxy \citep{ven02,mil04}. This suggests that the centre of the proto-cluster is not necessarily to the southeast of the radio galaxy, as suggested by the distribution of \Lya\ emitters. 
None of the 14 spectroscopically confirmed \Lya\ emitters in the MAMBO field were detected at 1.2~mm or 1.4~GHz. Similarly, none of the MAMBO sources show excess \Lya\ emission (if they are within the $4.075<z<4.123$ coverage of the narrow-band \Lya\ filter). A possible explanation would be that their large amounts of dust, as traced by the 1.2~mm emission may have quenched the \Lya\ emission \citep[\eg][]{kun98}, putting them below the narrow-band excess cutoff (EW$_{\rm rest}>$15~\AA). This apparent absence of overlap between both populations shows the importance of using multiple wavelength techniques to obtain a more complete picture of the proto-cluster. \citet{cha00} and \citet{webb03} also report a $<$20\% overlap between SMGs and LBGs, with a few notable exceptions \citep[\eg][]{cha02}. However, \citet{webb03} do find a large cross-clustering correlation amplitude between both populations in the largest of their 3 fields. We do not detect enough MAMBO sources in the TN~J1338$-$1942 field to perform such a cross-correlation analysis, but our detection of a statistical overdensity of MAMBO sources does suggest some relation to the confirmed overdensity of \Lya\ emitters \citep{ven02} and LBGs \citep{mil04} in the same field. 

The faintness of the \Lya\ in our objects also contrasts the published spectra of SMGs, which often show bright \Lya\ lines \citep{cha03}. However, these SMGs are on average at lower redshifts $\langle z \rangle \sim 2.4$ \citep{cha03}, while our MAMBO sources are potential members of the $z$=4.1 proto-cluster. Although we only have $K-$band information for 3 sources, which are all relatively bright class-II ($I-K\simlt 5$, $K\simlt $21) or class-I ($I-K\simgt 5$) sources \citep{ivi00,sma02a}, 3 of the 4 brightest MAMBO sources may well be very faint class-0 ($K\simgt 21$) sources which are much harder to obtain redshifts from \citep[\eg][]{dan02,fra04}, either because they are more highly obscured, or because they are at higher redshifts.
Our success in finding higher redshift objects may have been helped by the use of a longer selection wavelength (1.2~mm instead of 850~$\mu$m). 
Indeed, based on the low ratio of 850~$\mu$m to 1.2~mm flux, several authors argue that a significant fraction of mm galaxies may be at $z>3$ \citep[\eg][]{eal03,are03}. Most MAMBO sources in the NTT deep field \citep{dan02} are also class-0 sources, supporting the trend that 1.2~mm selected sources appear to be fainter (and hence maybe at higher redshift) than 850~$\mu$m selected sources. 

The detection of X-ray emission in several bright submm sources within the proto-clusters surrounding three other HzRGs \citep{sma03b} suggests that AGN may be present within those sources. Half of the spectra presented by \citet{cha03} also show type-II AGN lines, but we failed to detect these in our very deep VLT spectroscopy of M01 and M05. Hence, we have no indication that AGN are present in our MAMBO sources, but their emission lines may well have been obscured \citep[\eg][]{reu03}. Assuming the MAMBO sources are at $z$=4.1, and dust parameters $T_{\rm d}=$50~K and $\beta$=1.5 \citep{ben99}, we derive $L_{\rm FIR} \approx 2\times 10^{13}{\rm L_{\odot}} \times S_{\rm 1.2 mm}/{\rm mJy}$ in the range 4--12$\times 10^{13} {\rm L_{\odot}}$, implying star formation rates of 4000 to 12000~M$_{\odot}$yr$^{-1}$, if the dust is entirely heated by star formation \citep[\eg][]{omo01,deb03}. For M08, which is possibly at $z$=1.18, we derive $L_{\rm FIR} \approx 1.6\times 10^{13}{\rm L_{\odot}}$ and a star formation rate of 1600~M$_{\odot}$yr$^{-1}$. Such high star formation rates have been reported for HzRGs \citep{arc01,reu04}, but our MAMBO map indicates that they may also occur out to distances as far as 2~Mpc from the central radio galaxy. If the MAMBO sources are really at $z$=4.1, this would suggest that these proto-clusters have multiple, possibly aligned \citep{pen02} concentrations. However, the VLT narrow-band \Lya\ image shows that none of the mm sources seen in the TN~J1338$-$1942 proto-cluster have huge \Lya\ haloes, like those seen around the central HzRG \citep[\eg][]{ven02,reu03}. The HzRG therefore appears the best candidates to evolve into the present-day giant elliptical.

\section{Conclusions}
We summarize the results from our multi-wavelength observations of the $z$=4.1 proto-cluster surrounding TN~J1338$-$1942 as follows:

$\bullet$ We detect 10 candidate mm sources with peak fluxes having S/N$>$3 in our MAMBO map. Of these, at least eight sources with $S_{\rm 1.2 mm}>$1.3~mJy have possible radio and/or optical/near-IR counterparts, and 5 are confirmed at S/N$>$4 with pointed SCUBA submm photometry. Three sources have $S_{\rm 1.2 mm}>$4.0~mJy, while comparing with source counts from blank field surveys, we expect to find only 1 such source in the unassociated field population.

$\bullet$ The radio-to-submm and mm-submm photometric redshift estimates do not provide strong constraints on the possible redshifts of the MAMBO sources, although for 9 of the 10 sources, they are consistent with $z$=4.1 within the uncertainties.

$\bullet$ One of the faint MAMBO sources is possibly related to an ERO with $R-K$=6.1, which has a likely spectroscopic redshift of $z$=1.18, implying a star formation rate up to 1600~M$_{\odot}$yr$^{-1}$.

$\bullet$ None of the 14 spectroscopically confirmed \Lya\ emitters show detectable millimetre emission, and supposing they are at $z$=4.1, none of the 1.2~mm sources show an excess \Lya\ emission, indicating no apparent overlap between both populations.

$\bullet$ The mean star formation rate of the 14 spectroscopically confirmed \Lya\ emitters, as derived from the deep \Lya\ imaging is two orders of magnitudes lower than the upper limits derived from the stacked VLA and MAMBO maps, illustrating that the radio and mm maps probe much more actively star-forming galaxies.

Our multi-wavelength observations of this $z$=4.1 proto-cluster suggest that the \Lya\ excess technique does not detect the massive starburst companions within 2~Mpc. Their high SFR could be triggered by interactions between different proto-cluster members, as indicated by the diffuse nature of the optical identifications of M05 and M02.

\begin{acknowledgements}
We thank Thomas Greve for useful discussions.
IRAM is supported by INSU/CNRS (France), MPG (Germany) and IGN (Spain). 
The National Radio Astronomy Observatory (NRAO) is operated by Associated Universities, Inc., under a cooperative agreement with the National Science Foundation. 
The JCMT is operated by JAC, Hilo, on behalf of the parent organizations of the Particle Physics ans Astronomy Research Council in the UK, the National Research Council in Canada, and the Scientific Research Organization of the Netherlands.
This work was supported by a Marie Curie Fellowship of the European Community programme 'Improving Human Research Potential and the Socio-Economic Knowledge Base' under contract number HPMF-CT-2000-00721, and by the European RTN programme ``The Physics of the Intergalactic Medium''.
The work by MR and WvB at IGPP/LLNL was performed under the auspices of the U.S. Department of Energy, National Nuclear Security Administration by the University of California, Lawrence Livermore National Laboratory under contract No. W-7405-Eng-48. This work was carried out in the context of EARA, the European Association for Research in Astronomy.
\end{acknowledgements}

\end{document}